\documentclass[twocolumn,11pt]{article}
\usepackage{times,amsmath,amsthm,graphicx,float,hyperref,xcolor,booktabs,longtable,bm}
\newtheorem{definition}{Definition}

\newtheorem{example}[definition]{Example}
\newcommand{\R}{{\rm I}\!{\rm R}} % the set of real numbers
\newcommand{\argmin}{\mathop{\mathrm{argmin}}\limits}
\newcommand{\minimo}{\mathop{\mathrm{min}}\limits}
\newcommand{\maximo}{\mathop{\mathrm{max}}\limits}

\newcommand{\ds}{\displaystyle}
\begin{document}
\global\def\refname{{\normalsize \it References:}}
\baselineskip 12.5pt
%
%
% TITLE, AUTHOR, ABSTRACT, KEYWORDS
%
\title{\LARGE \bf Optimal estimation of thermal diffusivity\\ in an energy transfer problem}

\date{}

\author{\hspace*{-10pt}
\begin{minipage}[t]{2.7in} \normalsize \baselineskip 12.5pt
\centerline{GUILLERMO F. UMBRICHT}
\centerline{Universidad Nacional de Gral.~Sarmiento}
\centerline{Instituto de Ciencias}
\centerline{Instituto del Desarrollo Humano}
\centerline{J. M. Guti\'errez 1150}
\centerline{Los Polvorines}
\centerline{ARGENTINA}
\end{minipage} \kern 0in
\begin{minipage}[t]{2.7in} \normalsize \baselineskip 12.5pt
\centerline{DIANA RUBIO}
\centerline{Universidad Nacional de San Mart\'in}
\centerline{Escuela de Ciencia y Tecnolog\'ia}
\centerline{Centro de Matem\'atica Aplicada}
\centerline{ITECA (CONICET-UNSAM)}
\centerline{25 de Mayo y Francia, San Mart\'in}
\centerline{ARGENTINA}
\end{minipage}
%
% If you are three authors then you can use three mini--pages
% instead of two. Their horizontal size must be less than 2.7in
% indicated above. It can be e.g. 2.3in. However, you must pay
% attention that you do not exceed the total width of the text.
%
\\ \\ \hspace*{-10pt}
\begin{minipage}[b]{6.9in} \normalsize
\baselineskip 12.5pt {\it Abstract:}
% The text of the abstract follows
%/
This work focuses on determining the coefficient of thermal diffusivity in a one-dimensional heat transfer process along a homogeneous and isotropic bar, embedded in a moving fluid with heat generation. A first type (Dirichlet) condition is imposed on one boundary and a third type (Robin) condition is considered at the other one.
The parameter is estimated by minimizing the squared errors where noisy observations are numerically simulated at different positions and instants.  The results are evaluated by means of the relative errors for different levels of noise. In order to enhance the estimation performance, an optimal design technique is chosen to select the most informative data. Finally, the improvement of the estimate is discussed when an optimal design is used.
\\ [4mm] {\it Key--Words:}
Parabolic equation, heat transfer, parameter estimation, mathematical modeling.
\end{minipage}}
\vspace{-10pt}

\maketitle

\thispagestyle{empty} \pagestyle{empty}
% numbers of pages are supplemented by the editor
%
% THE BEGINNING OF THE TEXT
%
\section{Introduction}
\label{S1} \vspace{-4pt}

This work deals with the inverse problem consisting in the determination of the coefficient of thermal diffusivity in a heat transfer problem with convection and thermal dissipation by lateral heat flow. 
The estimation of this parameter in heat transfer problems  has diverse applications, for instance the design of an optimal control system in a thermal process, among others. There are few publications dedicated to the direct identification of this thermal parameter, among them \cite{Umbricht15}. On the other hand, a number of articles are devoted to obtain the thermal diffusivity by means of other thermal coefficients.
When the thermal diffusivity is assumed constant, it can be obtained from the simultaneous estimation of two other thermal coefficients: the thermal conductivity along with  specific heat or heat capacity. This is due to the fact that they have a determining influence on the distribution of temperatures and heat flow densities during transient heating or cooling processes.
Several works have been dedicated to the identification of these parameters from temperature data (see, for example,  \cite{Chen98}, \cite{Dantas96}, \cite{Huang91},  \cite{Sawaf95}, \cite{Yang00}). More recently, interesting works on the simultaneous estimation of the thermal conductivity and specific heat coefficients were studied in \cite{Kim02, Kim02b}.

Thermal energy transfer processes have been widely studied during the last decades from the different branches of science and technology, \cite{Pacheco14,Rao2018,Stewart13}.
These types of problems are still being addressed a it can be seen in \cite{Balaji2021,Dincer2021,Stewart21}.

Recent research shows, in different contexts, the importance of studying the estimation of thermal parameters in heat transfer processes. Several applications as well as the development of new analytical and numerical techniques indicate the relevance of 
the determination of these parameters. See, as an example, \cite{Chen16}, \cite{Wang19}.
The parameters that are most frequently estimated, under different assumptions and with different techniques, are: thermal conductivity \cite{Pourrajab21}, \cite{Umbricht21b}, thermal diffusivity \cite{Beck21}, \cite{Umbricht15}, sources of heat generation \cite{Umbricht19}, \cite{Umbricht21}, \cite{Zhao14} and the coefficient of heat transfer by convection \cite{Rashad14}, \cite{Umbricht20}.

In this work the estimation of the thermal diffusivity coefficient is conducted based on numerically simulated  noisy data at few positions and instants of time. 
Firstly, the parameter is estimated by minimizing the squared errors, including a numerical  analysis on the number of observations data required for the estimation. Afterward, a numerical sensitivity analysis is performed in order to improve the identification using an optimal design technique that provides a criterion for selecting the most informative data. Numerical experiments are made and relative errors are calculated to evaluate the performance of the determination procedure. Finally, a discussion on the advantage of using an optimal design to select observation data for estimation is included.

\section{Mathematical formulation of the problem}
\label{Presentation} \vspace{-4pt}
In this work we focus on the determination of the coefficient of thermal diffusivity of a material involved in a one-dimensional transport or transfer problem of thermal energy. It is assumed that the process occurs along a homogeneous bar of length $ L \, [m] $ and diameter $ d \, [m] $ embedded in a fluid that stays at temperature $ T_a \, [^{\circ}C]$ and moves with constant velocity at a speed $\beta \, [m / s]$.
It is also assumed that  bar is isotropic, so that the coefficient of thermal diffusivity $ \alpha^2 \, [m^2 / s] $ is constant. At the left boundary of the bar,  a constant temperature $ F \, [^{\circ}C] $ is imposed while the other end is left free in contact with the fluid allowing dissipation by convection. In addition, it is considered that there is a heat exchange  between the lateral surface of the bar and the fluid at a constant rate $\nu \, [1 / s]$.
 Finally, time-invariant external sources are considered, denoted by $ f(x) \, [^{\circ}C / s] $.

The thermal process is mathematically modeled using a parabolic equation with initial and boundary conditions. The aim of this work is the estimation of $ \alpha^2 $ of the following system for $ t> 0 $,
\begin{equation}
\label{Sistem}
\begin{cases} 
u_t(x,t)= D u(x,t) + f(x), \,\, 0<x<L,\\
u(0,t)=F,  \\
\kappa\, u_x(L,t)=-h\,(u(L,t)-T_a), 
\end{cases}
\end{equation}
with initial condition
\begin{equation}
\label{Initialu}
u(x,0)=T_a, \quad  0<x<L.
\end{equation}
The differential operator $D$ in \eqref{Sistem} is defined by
\begin{equation}
\label{Operatoru}
D u(x,t):=\alpha^2 u_{xx}(x,t)-\beta u_{x}(x,t) - \nu (u(x,t)-T_a), \end{equation}
 for $u \in C^2(0,L) \times C^1(0,+\infty)$ where $\kappa \, [W/m^{\circ}C]$ is the thermal conductivity coefficient of the bar material and $h \, [W/m^2\,{}^{\circ}C]$\cite{Umbricht20} is the heat transfer coefficient.

The estimation is carried out from simulated temperature  $u_\epsilon (x_i, t_j)$ at positions $x_i$ for $i = 1, 2,\dots, n$  and instants $t_j$ with $j = 1, 2, \dots, m$,  that satisfy 
\begin{equation}
\label{Data_condition}
\left|u(x_i,t_j)-u_\epsilon(x_i,t_j)\right|<\epsilon, 
\end{equation}
where $\epsilon$ is a bound for the noise in the data, or the measurement error, and $u(x_i,t_j) $ represent the exact temperature at $(x_i,t_j)$. 

\section{Estimation of thermal diffusivity}
\label{Estimation}\vspace{-4pt}
5
A least square method is applied for the estimation of the thermal diffusivity coefficient $\alpha^2$ that appears in the differential operator $D$ defined in \eqref{Operatoru},
%, for the process described by the system \eqref{Sistem}-\eqref{Data_condition}
 as it is described below.
% This optimization approach consists in  looking for the parameter that minimize the squared %errors between the simulated temperature and the respective data.

Let the functional $J:\Theta \subset \R^{+} \to  \R^{+}$  be
\begin{equation}
\label{Functional_J}
J(\alpha^2):=\ds \sum_{i=1}^{n}\sum_{j=1}^{m}\left|u(x_i,t_j,\alpha^2)-u_\epsilon(x_i,t_j)\right|^2,
\end{equation}
where $\Theta$ is the set of admissible values for the parameter to be estimated.
We look for the parameter value $\widehat{\alpha^2}$ that minimizes the functional \eqref{Functional_J}, that is,
\begin{equation}
\label{Theta_optimo}
\widehat{\alpha^2}=\argmin_{\alpha^2 \in \Theta} J(\alpha^2), 
\end{equation}

In the particular case of the estimation of the thermal diffusivity coefficient for metallic materials, the set of admissible values could be defined as $\Theta:= (5 \times 10^{-6},5 \times 10^{-4})$ (see Table \ref{thermal_prop}). When other types of materials are considered, the set $\Theta$  has to be defined accordingly.
\begin{table}[H]
\begin{center}
\begin{tabular}{lccc}
\hline
Material (Symbol)	& $k [W/m ^\circ C]$ &	$\alpha^2 \times 10^4 
[m^2/s]$ \\ 
\hline
Lead	(Pb) & 35 & 0.23673\\
Nickel	(Ni) & 70 & 0.22660\\
Iron	(Fe) & 73 & 0.20451\\
%Magnesium (Mg) & 156 & 0.88300\\
Aluminum (Al) & 204 & 0.84010\\
Copper (Cu) & 386 & 1.12530\\
Silver (Ag) & 419 & 1.70140\\
\hline
\end{tabular}
\end{center}
\vspace{-0.5cm}
\caption{\label{thermal_prop} Thermal properties of different materials}
\end{table}
The existence of a unique global solution to problem \eqref{Functional_J}-\eqref{Theta_optimo} is guaranteed by the form of the functional $J$ \cite{Dennis83}. 
The optimization process is performed numerically for different data errors and different number of measurements.% at different time instants Some examples are included in the following section.

For the numerical examples, we use a finite difference scheme of second order centered in space and forward in time,  on a uniformly spaced grid.  This explicit method is convergent and stable for $\alpha^2 < \frac{h_x^2}{2 h_t}$  (\cite{Morton05},\cite{Umbricht15})  where   $h_x=0.01 \, m$ and $h_t=0.1 \, s$ are the spatial and temporal discretization steps considered for this work.
Random noise is added to numerical data in order to simulate experimental measurements. For the performance analysis, mean values of the relative errors for $N=1000$ runs are calculated. 
\begin{example}
\label{Ex1}
The determination of the thermal diffusivity for the aluminum and copper  is considered under the following set up:
$L=1 \, m$; $ F=100 \, ^{\circ}C$; $T_a=25 \,^{\circ}C$; $h=10 \, W/(m^{2}\,^{\circ}C)$;
$f(x)=-\dfrac{1}{50}x(x-1) \, ^{\circ} C/s$; $\beta=0.01 \, m/s$ and $\nu=0.0001 \, 1/s$.
\end{example}
%
%The temperature distribution is computed numerically for a uniformly spaced grid using an explicit  finite difference scheme of second order centered in space and forward in time. Afterwards, random noise is added in order to simulate experimental measurement data. 
%For the examples presented in this work, the temperature observations were considered with the %following noise levels:
The relative errors  between the estimated and the actual value  are calculated by
\begin{equation}
\label{Err}
Err=\frac{|\widehat{\alpha^2} - \alpha^2|}{ \alpha^2},
\end{equation}
$Err_{\rm{Al}}$, $Err_{\rm{Cu}}$ denote the relative errors of the estimation of the aluminum and copper thermal diffusivities, respectively.

Firstly, temperature data are simulated at three different positions 
$(x=0,x=L/2,x=L)$, at two instants: $t=12 \, h$ and $t=24 \, h$, considering the following noise levels :
 $\epsilon=0.5\,^\circ C;$  $ 1.0\,^\circ C;$  $1.5\,^\circ C;$  $2.0\,^\circ C;$ $2.5\, ^\circ C$. 
\begin{table}[H]
\begin{center}
\begin{tabular}{ccc} 
\toprule
   $\epsilon \,\left[^\circ C\right]  $   &  \qquad  $Err_{\rm{Al}}$  &   \qquad  $Err_{\rm{Cu}}$  \\\hline
{\begin{tabular}{c} 
 0.5 \\
 1.0 \\
 1.5 \\
 2.0 \\
 2.5  
\end{tabular}}& \qquad
{\begin{tabular}{c} 
 0.0025 \\
 0.0035 \\
 0.0053 \\
 0.0086 \\
 0.0098 
\end{tabular}}\qquad  & \qquad
{\begin{tabular}{c}    
 0.0007 \\
 0.0016 \\
 0.0043 \\
 0.0061 \\
 0.0081 
\end{tabular}} \\\bottomrule
\end{tabular} 
\end{center}
\vspace{-0.5cm} 
\caption{Example \ref{Ex1}: Relative estimation errors for Al ($Err_{\rm{Al}}$ ) and for Cu ($Err_{\rm{Cu}}$ ) taking $n=3$ equally spaced data $(x=0,x=L/2,x=L)$ at $t= 12\,h$ for different noise levels}
\label{table1}
\end{table}
\begin{table}[H]
\vspace{-0.5cm}
\begin{center}
\begin{tabular}{ccc} 
\toprule
   $\epsilon \,\left[^\circ C\right]  $   &  \qquad  $Err_{\rm{Al}}$  &   \qquad  $Err_{\rm{Cu}}$  \\\hline
{\begin{tabular}{c} 
 0.5 \\
 1.0 \\
 1.5 \\
 2.0 \\
 2.5  
\end{tabular}} &\qquad
{\begin{tabular}{c} 
 0.0003 \\
 0.0022 \\
 0.0059 \\
 0.0080 \\
 0.0122 
\end{tabular}} \qquad  & \qquad
{\begin{tabular}{c}    
  0.0013 \\
  0.0030 \\
  0.0066 \\
  0.0078 \\
  0.0119 
\end{tabular}}\\\bottomrule
\end{tabular} 
\end{center}
\vspace{-0.5cm} 
\caption{Example \ref{Ex1}: Relative estimation errors  for Al ($Err_{\rm{Al}}$ ) and for Cu ($Err_{\rm{Cu}}$ )  taking $n=3$ equally spaced data $(x=0,x=L/2,x=L)$ at $t= 24\,h$ for different noise levels.}
\label{table2}
\end{table}
Table \ref{table1} shows the relative errors for the estimation of  the thermal diffusivity coefficient for aluminum and for copper.  It can be observed that  
 $Err_{\rm{Al}}\in (0.0025, 0.0098) $, 
 $Err_{\rm{Cu}} \in  (0.0007 , 0.0081)$. In other words, the relative errors for the aluminum bar vary in the interval $0.25 \% $ - $0.98 \% $ while the corresponding ones for copper vary in $0.07 \% $ - $0.81 \% $ . In both cases, the relative  errors remain smaller than $1 \%$ and increase as  the  data error increases.

The results obtained at $t=24\,h$ are shown in Table \ref{table2}. In this case the relative errors reach almost $1,25 \%$ for a data error of $2.5\, ^\circ C$, which implies that  a big measurement error leads to big relative errors. Note that the errors of the same order are obtained in both cases for both materials, i.e., no significant differences are found regarding the materials.

The dependency  on the number of measurement data, $n$,  is also numerically investigated. 
\begin{table}[H]
\begin{center}
\begin{tabular}{ccc} 
\toprule
   $n$   &  \qquad  $Err_{\rm{Al}}$  &   \qquad  $Err_{\rm{Cu}}$  \\\hline
{\begin{tabular}{l} 
 \,1 \\
 \,3 \\
 \,5 \\
 10 
\end{tabular}}&  \qquad 
{\begin{tabular}{c} 
  0.0055 \\
  0.0017 \\
  0.0011 \\
  0.0037   
\end{tabular}}  \qquad  &\qquad  
{\begin{tabular}{c}    
  0.0024 \\
  0.0018 \\
  0.0010 \\
  0.0021   
\end{tabular}} \\\bottomrule
\end{tabular}
\end{center}
\vspace{-0.5cm} 
\caption{Example \ref{Ex1}: Relative estimation errors taking $n$ equally spaced data at $t= 12\,h$ for $\epsilon=1 \, ^\circ C$.}
\label{table3}
\end{table}
\vspace{-0.5cm} 
\begin{table}[H]
\begin{center}
\begin{tabular}{ccc} 
\toprule
   $n$   &  \qquad  $Err_{\rm{Al}}$  &   \qquad  $Err_{\rm{Cu}}$  \\\hline
{\begin{tabular}{c} 
 \,1 \\
 \,3 \\
 \,5 \\
 10 
\end{tabular}}&  \qquad 
{\begin{tabular}{c} 
  0.0568 \\
  0.0010 \\
  0.0009 \\
  0.0001 
\end{tabular}} \qquad   &\qquad   
{\begin{tabular}{c}    
  0.0956 \\
  0.0030 \\
  0.0021 \\
  0.0013 
\end{tabular}} \\\bottomrule
\end{tabular} 
\end{center}
\vspace{-0.5cm} 
\caption{Example \ref{Ex1}: Relative estimation errors  taking $n$ equally spaced data at $t= 24\,h$ for $\epsilon=1 \, ^\circ C$.}
\label{table4}
\end{table}
Tables \ref{table3} and \ref{table4} show the results for $n=1,3,5,10$ when considering a noise level of $\epsilon=1,0 \, ^\circ C$ which is a reasonable measurement error for the range of temperature that we are considered here.
As before, the results correspond to an aluminum bar and a copper bar.
In  Table \ref{table3} are shown the relative errors for the estimation of the thermal diffusivity for both materials. The results indicate that all of them are of  the same order, where the smallest one is reached for $n=5$, being $0.11\, \%$ for Al and $0.10 \,\%$ for Cu.
Table \ref{table4} shows the corresponding relative errors for numerical observations at $t=24\,h$. For $n=1$, the greatest errors are obtained. As $n$ increases, an improvement is observed reaching the best performance at $n=10$ with relative errors of $0.01\,\%$ and $0.13\, \%$ for Al and Cu, respectively.  
\section{Sensitivity analysis}
\label{Sensitivity_analysis}\vspace{-4pt}

The main objective of the sensitivity analysis is to quantitatively determine the change in the behavior of a system when the values of a set of its parameters is modified. Here, 
the sensitivity function \cite{Banks07,Chatt88,Kittas97,Thomaseth99,Umbricht15} is used  to analyze the local influence of the thermal diffusivity coefficient, $\alpha^2$, in the calculation of the temperature, $u(x,t)$. Later, this analysis will be used to determine where and when the most informative data can be obtained. 

The sensitivity of  $u(x,t)$ with respect to $\alpha^2$ is defined as 
\begin{equation}
\label{Funciones_Sensi}
S(x,t)=S_{u}^{\alpha^2}=\dfrac{\partial u(x,t)}{\partial \alpha^2}.
\end{equation}

\begin{figure}[H]
\begin{center}
\includegraphics[width=0.4\textwidth]{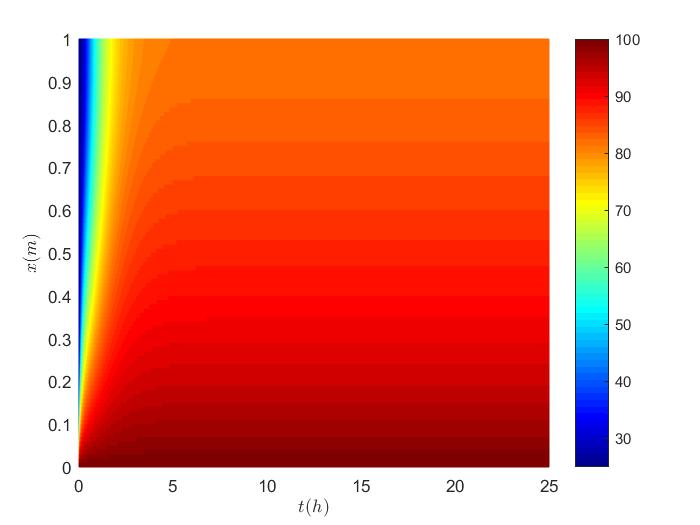}\vspace{-0.13cm}
%\end{center}
%\vspace{-0.5cm}
%\caption{Example \ref{Ex1}: Temperature function for a aluminum bar.}
%\vspace{-0.8cm}
%\label{Fig5}
%\end{figure}
%
%\begin{figure}[H]
%\begin{center}
\includegraphics[width=0.4\textwidth]{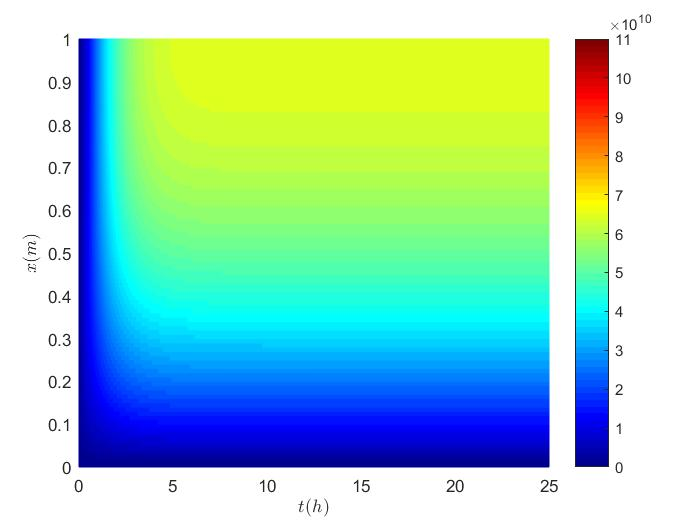}
\end{center}
\vspace{-0.8cm}
%\caption{Example \ref{Ex1}: Absolute sensitivity function for a aluminum bar.}
\caption{Example \ref{Ex2}: Temperature (top) and Absolute sensitivity (bottom)  functions for a aluminum bar.}
%\vspace{-0.8cm}
\label{Fig1}
\end{figure}

\begin{figure}[H]
\begin{center}
\includegraphics[width=0.4\textwidth]{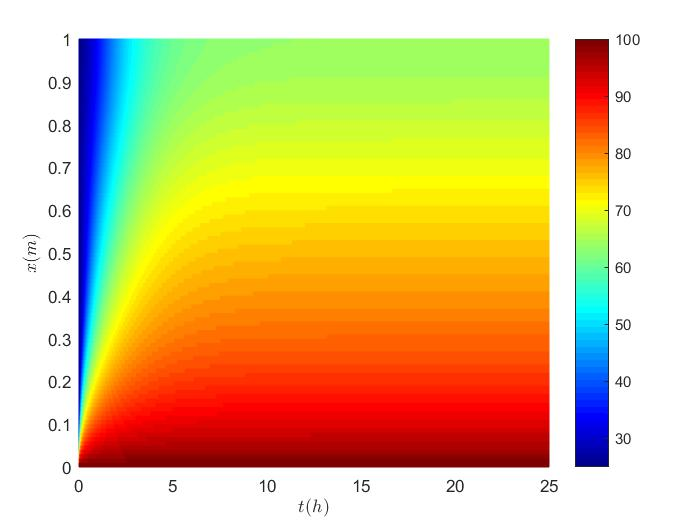}\vspace{-0.13cm}\\
%\end{center}
%\vspace{-0.5cm}
%\caption{Example \ref{Ex1}: Temperature function for a iron bar.}
%\vspace{-0.8cm}
%\label{Fig5}
%\end{figure}
%\begin{figure}[H]
%\begin{center}
\includegraphics[width=0.4\textwidth]{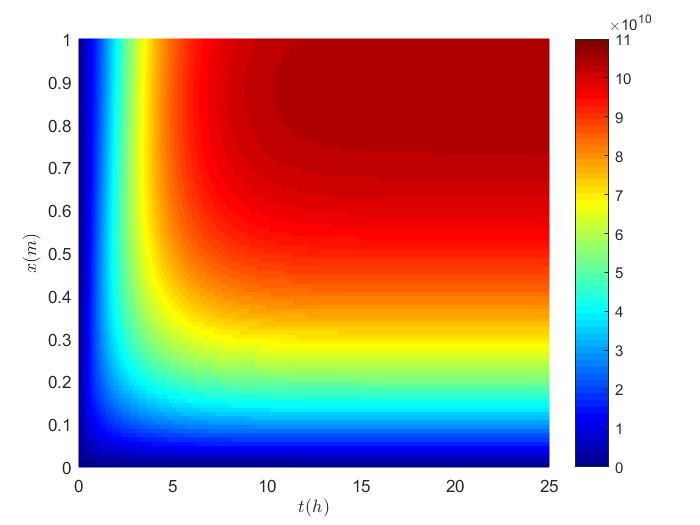}
\end{center}
%\vspace{-0.8cm} 
\caption{Example \ref{Ex2}: Temperature (top) and absolute sensitivity (bottom) functions for an iron bar.}
%\vspace{-0.8cm} 
\label{Fig2}
\end{figure}

The system  \eqref{Sistem}-\eqref{Operatoru} is derived with respect to $\alpha^2$, for $t>0$. Since $\frac{\partial f(t)}{\partial \alpha^2} =0$, it results
\begin{equation}
\label{System_derivado}
\begin{cases} 
\dfrac{\partial u_t(x,t)}{\partial \alpha^2}= \dfrac{\partial Du(x,t)}{\partial \alpha^2}, \,\, 0<x<L,\vspace*{0.25cm}\\
\dfrac{\partial u(0,t)}{\partial \alpha^2}=0,  \vspace*{0.25cm}\\
\kappa \dfrac{\partial \, u_x(L,t)}{\partial \alpha^2}=-h\,\dfrac{\partial u(L,t)}{\partial \alpha^2}, 
\end{cases}
\end{equation}
with initial condition
\begin{equation}
\dfrac{\partial u(x,0)}{\partial \alpha^2}=0, \quad  0<x<L.
\end{equation}
Properties of partial derivatives and the definition  \eqref{Funciones_Sensi} allow us to write
\begin{eqnarray} 
   \label{Sxt}
   \dfrac{\partial u_t(x,t)}{\partial \alpha^2}&=& \dfrac{\partial^2 u(x,t)}{\partial t \, \partial \alpha^2}=
  \dfrac{\partial S(x,t)}{\partial t}\\
  \label{Sxx}
  \dfrac{\partial u_{xx}(x,t)}{\partial \alpha^2}&=&S_{xx}(x,t), \\ 
  \label{Sx}
  \dfrac{\partial u_{x}(x,t)}{\partial \alpha^2}&=&S_x(x,t)
   \end{eqnarray} 
 leading to
\begin{equation}
\begin{split} 
\dfrac{\partial u_t(x,t)}{\partial \alpha^2} & =  u_{xx}(x,t)+\alpha^2 \dfrac{\partial u_{xx}(x,t)}{\partial \alpha^2} \\
& -\beta \dfrac{\partial u_{x}(x,t)}{\partial \alpha^2} - \nu \dfrac{\partial u (x,t)}{\partial \alpha^2}
 \end{split}
   \end{equation}
   or, equivalently, by \eqref{Sxx}-\eqref{Sx} 
   \begin{equation}
\label{partialD}
\begin{split} 
\dfrac{\partial u_t(x,t)}{\partial \alpha^2} 
   & = u_{xx}(x,t)+\alpha^2  S_{xx}(x,t)\\
   &-\beta S_{x}(x,t) - \nu S(x,t).
   \end{split}
   \end{equation}
Since  $u \in C^2(0,L) \times C^1(0,+\infty)$, we have that $S \in C^2(0,L) \times C^1(0,+\infty)$ and, by the definition of the operator $D$ given in  \eqref{Operatoru}, it follows that
 \begin{equation}
 \label{DS}
\dfrac{\partial Du(x,t)}{\partial \alpha^2}
   = u_{xx}(x,t) + DS(x,t) - \nu T_a.
\end{equation}
Equations \eqref{System_derivado}-\eqref{DS} lead to a system for the sensitivity function that involves  $u(x,t)$, that is, a coupled system for $u(x,t)$ and  $S(x,t)$ in  $(0,L) \times(0,+\infty)$ is obtained, 
\begin{equation}
\label{Sensi}
\begin{cases} 
u_t(x,t)= D u(x,t) + f(x),&  \\
S_t(x,t)= DS(x,t) + u_{xx}(x,t) - \nu T_a, & \\
\end{cases}
\end{equation}
with boundary conditions
\begin{equation}
\begin{cases} 
u(0,t)=F, &\\%\, & x=0 , \, t>0, \\
S(0,t)=0, &\\% \, & x=0 , \, t>0, \\
\kappa\, u_x(L,t)=-h\,(u(L,t)-T_a) , &\\%\, & x=L ,  t>0, \\
\kappa\, S_x(L,t)=-h\,S(L,t), & 
\end{cases}
\end{equation}
and initial conditions
\begin{equation}
\begin{cases}
\label{initialSu}
u(x,0)=T_a, &\\
S(x,0)=0. &
\end{cases}
\end{equation}
For the analysis, the absolute value of the sensitivity is considered, that is, $|S(x, t)|$, and we call it absolute sensitivity.
The system \eqref{Sensi}-\eqref{initialSu} 
is solved numerically by means of finite differences centered in space and forward in time. 
\begin{example}
\label{Ex2}
In this example, the absolute sensitivity is calculated for different materials under the same set up used in the Example \ref{Ex1}.
\end{example}
In Figures \ref{Fig1}-\ref{Fig2}, a color scale is used to plot the temperature (top) and the absolute sensitivity (bottom) as functions of  $x[m]$ and $t[s]$, for a aluminum bar and for an iron bar, respectively.

It can be seen that $|S(x,t)|$ achieves greater values  for the bar made of iron ( Figure \ref{Fig2}) than for the one made of aluminum ( Figure \ref{Fig1}).  
 For instance, at $x=0.5 \, m$ ($x=L/2)$ and $ t>15 \, h$,  the absolute sensitivity is approximately $6 \times 10^{10}$ (yellow color) in the case of the aluminum bar while it is about $9 \times 10^{10}$ (red color)  for the iron one.
This implies that  small changes in the thermal diffusivity coefficient yield greater changes in the  temperature function for iron than for aluminium. This  fact has a physical meaning, since it could be related to the property of the material to diffuse temperature.

In both Figures, greater values for the absolute sensitivity are achieved close to the right edge of the bar. On the other hand,  at each fixed $x$,  no significant changes are shown in Figure \ref{Fig1} for $t>5h$. The reason for this is that the temperature achieves its stationary value around $t=5\,h$ for the aluminum bar, as it can be seen in the plot. Observed that once the steady-state for the temperature is achieved, the sensitivity will only depend on the space variable (see Eq. \eqref{Funciones_Sensi}).
Analogously, in the case of the iron bar, the steady-state for both, the temperature and the absolute sensitivity, is achieved around $t=12\, h$, as it can be seen in  Figure \ref{Fig2}. 
%This is because  iron is a less diffusive material than aluminum (see Table \ref{thermal_prop}).

Now, the absolute sensitivity for different materials are compared, fixing one of the variables at a time, either $x$ or $t$. Specifically, we consider the functions $|S(L/2, t)|$ and $ |S(L, t)|$, and  functions  $|S(x, 5)|$ and $ |S(x, 15)|$.

\begin{figure}[H]
\begin{center}
\includegraphics[width=0.42\textwidth]{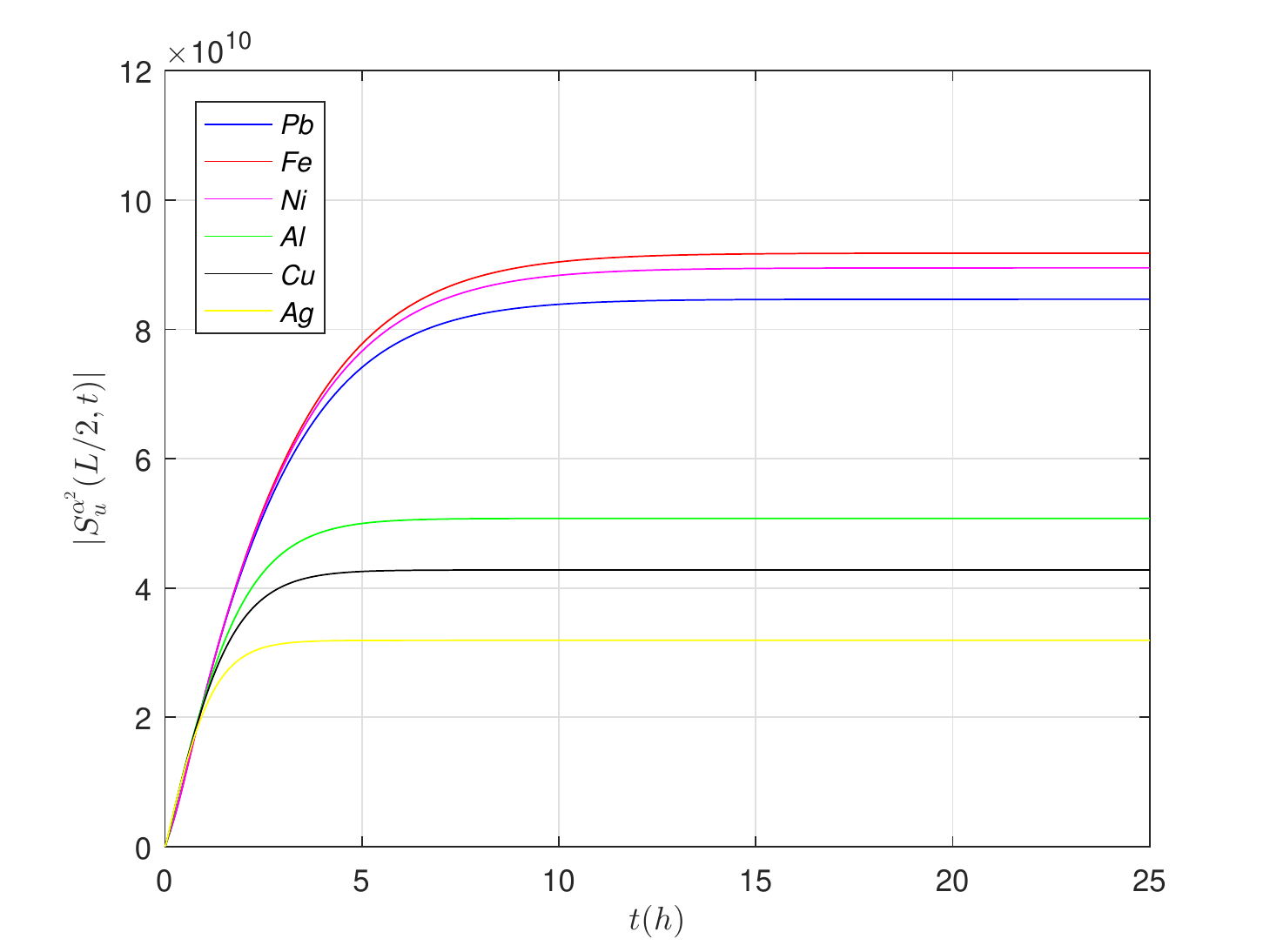}
\end{center}
\vspace{-0.5cm} 
\caption{Example \ref{Ex2}: Absolute sensitivity $|S(L/2, t)|$ for different materials at $(x=L/2)$.}
\label{Fig3}
\end{figure}

\begin{figure}[H]
\begin{center}
\includegraphics[width=0.42\textwidth]{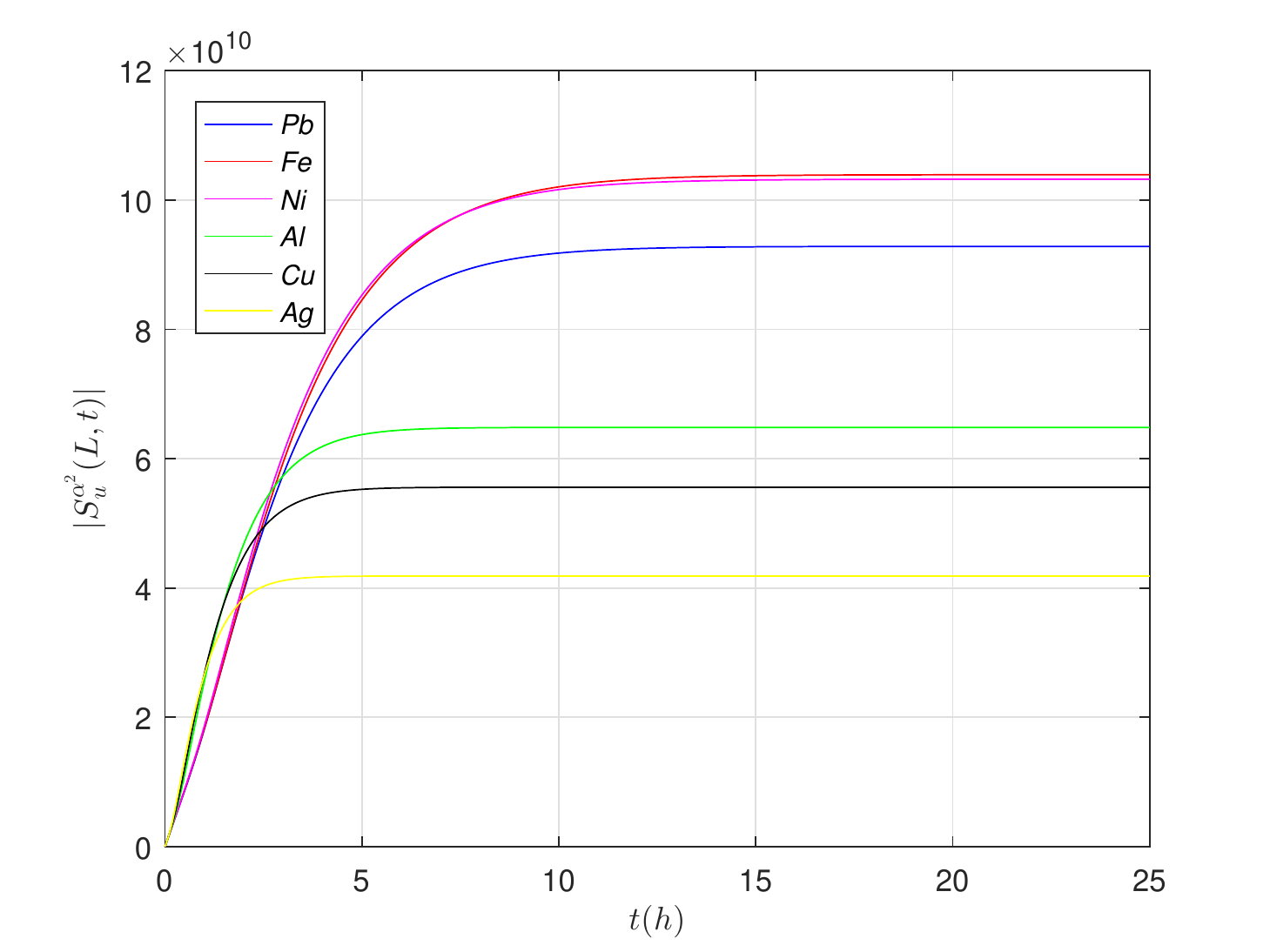}
\end{center}
\vspace{-0.5cm} 
\caption{Example \ref{Ex2}: Absolute sensitivity$|S(L, t)|$ for different materials at $(x=L)$.}
\label{Fig4}
\end{figure}

Figures \ref{Fig3}- \ref{Fig4} show the time profiles of the absolute sensitivity function evaluated at  the middle of the bar,  $(x = L / 2)$, and  at the right edge, $(x = L)$, respectively, for all materials included in Table \ref{thermal_prop}. 
Note that the absolute sensitivity increases  with time and that, in both cases, it reaches the steady state after approximately 15 h for all the materials considered. This is consistent with the observations made from Figures \ref{Fig1}-\ref{Fig2} for aluminum and iron.
Comparing the absolute sensitivity for the different materials, it can be seen that the smaller values correspond to more diffusive materials (Al,  Cu, Ag), as it was suggested from Figures \ref{Fig1}-\ref{Fig2}.

\begin{figure}[H]
\begin{center}
\includegraphics[width=0.42\textwidth]{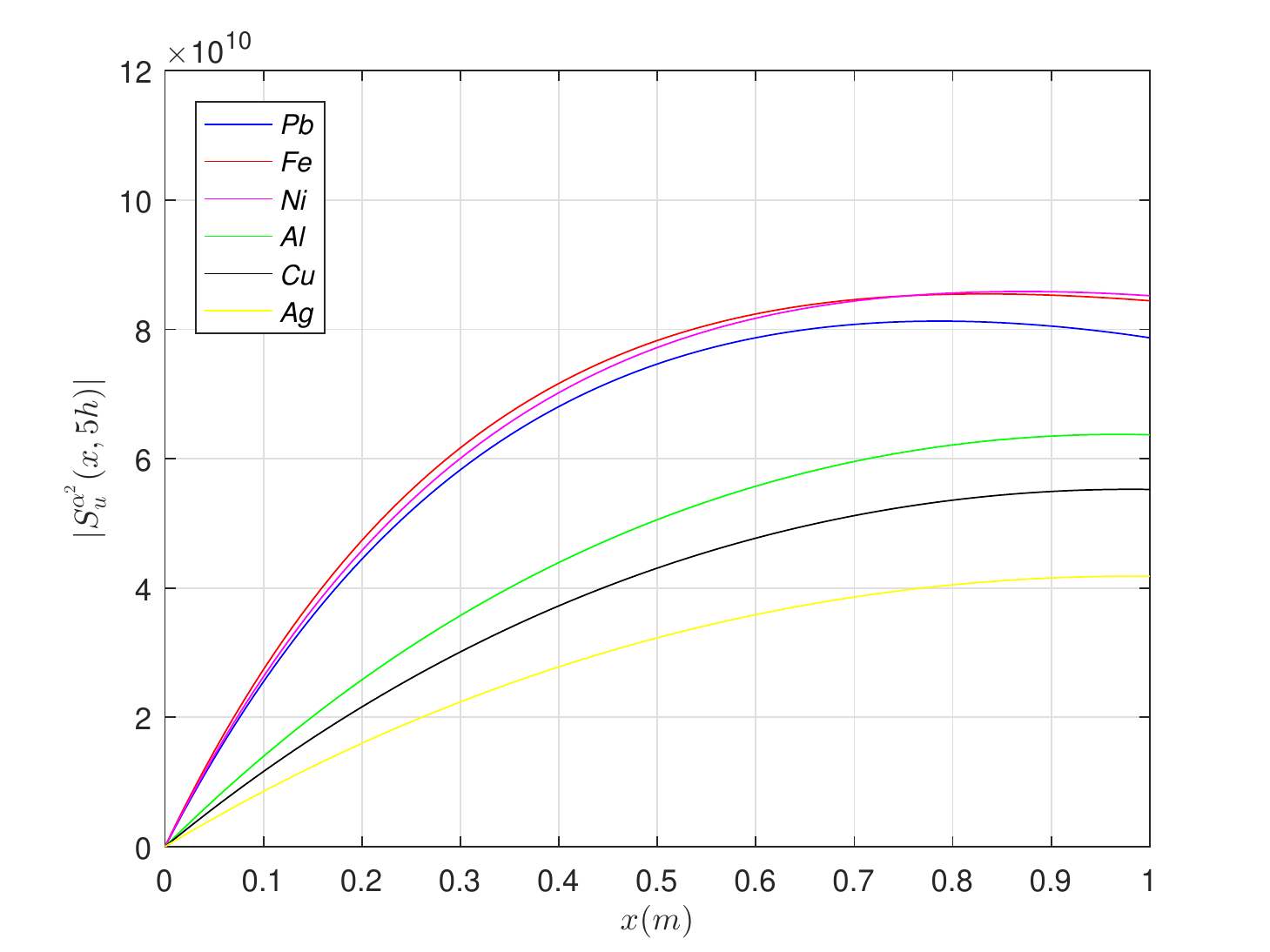}
\end{center}
\vspace{-0.5cm} 
\caption{Example \ref{Ex2}: Absolute sensitivity spatial profiles for different materials, at $t=5 \, h$.}
\vspace{-0.8cm}
\label{Fig5}
\end{figure}

\begin{figure}[H]
\begin{center}
\includegraphics[width=0.42\textwidth]{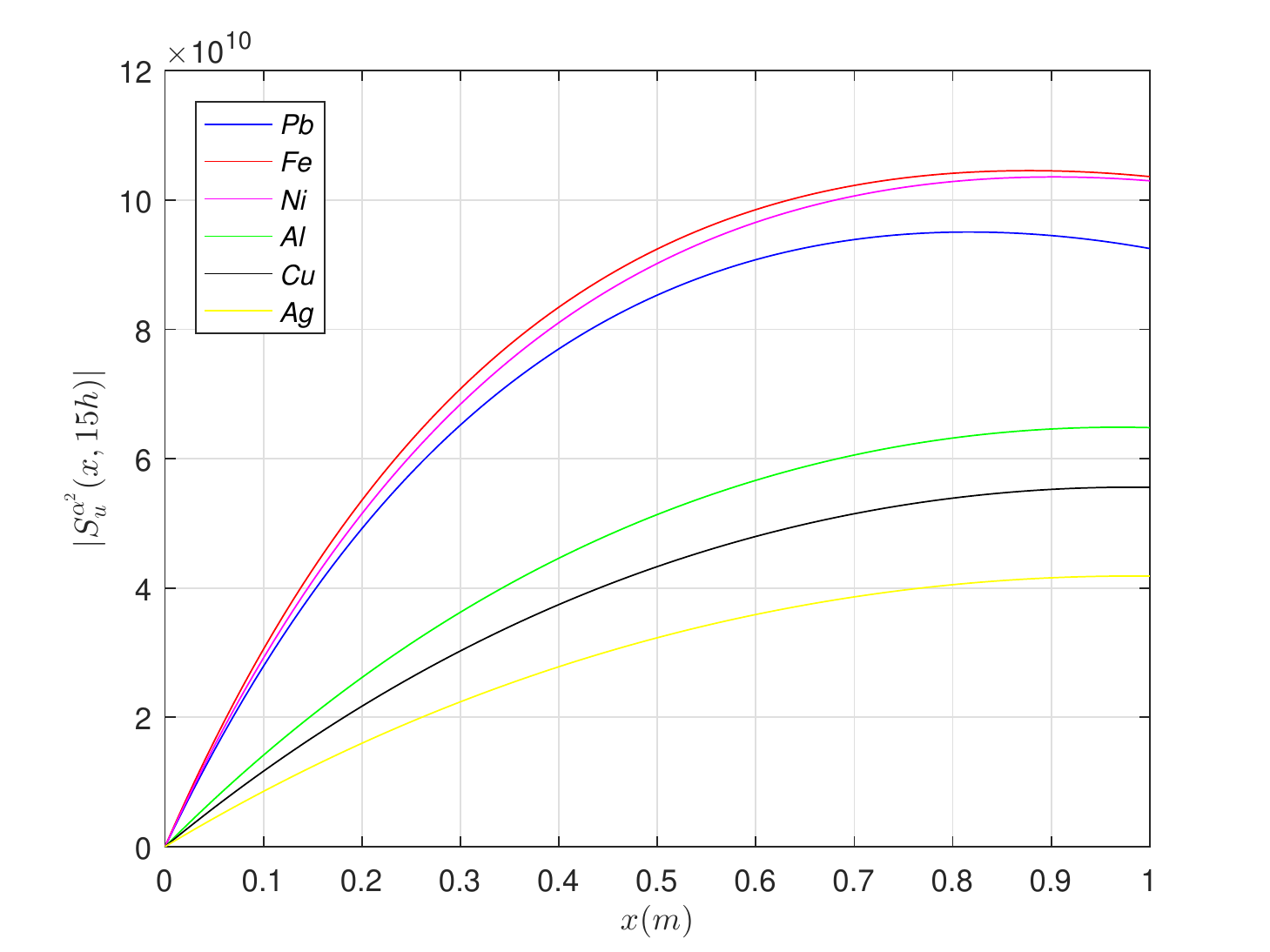}
\end{center}
\vspace{-0.8cm} 
\caption{Example \ref{Ex2}: Absolute sensitivity spatial profiles for different materials, at $t=15 \, h$.}
\label{Fig6}
\end{figure}

Figures \ref{Fig5}-\ref{Fig6} show the absolute sensitivity profiles at $t=5 \, h$ and $t=15 \, h$, respectively.
Again, as in the previous figures, it can be seen that  more diffusive materials lead to smaller absolute sensitivity  values. 
Comparing the spatial profiles at $t=5 \, h$ and $t=15 \, h$, no changes are observed for  Al, Cu, Ag (the more diffusive materials) while for  the other ones (Pb, Fe, Ni), big changes are shown. This is consistent with the previous observations.
On the other hand, the greater values are achieved close to $x=L$ in all cases.

\section{Optimal design}
\label{Optimal_design}\vspace{-4pt}

In this section an optimal design strategy is considered to determine locations and instants to measure data in order to obtain accurate estimations.
The theory of optimal design experiments \cite{Ucinski04,Van00,Vande00} and 
optimal location of measurement sensors \cite{Alana10,Banks13,Rensfelt08,Sumana09} have been extensively studied in different problems for several distributed parameter systems. In some of these works the optimal design theory is used for the identification of thermal parameters, as in \cite{Beck69, Umbricht15}.

Optimal design techniques usually  requires the minimization of a functional that involves the Fisher information matrix \cite{Banks10,Banks13,Banks15,Thomaseth99}. For  $n$  observations and $p$  parameters to be estimated, it is defined by 
\begin{equation}
\label{Fisher_n_parameters}
F_{ij}(\bm x, \bm t, \bm \theta)=\ds \sum_{k=1}^{n} S_{u}^{\theta_i}(x_k, t_k, \bm \theta) \, S_{u}^{\theta_j}(x_k, t_k, \bm \theta),
\end{equation}
where ${\bm x}=(x_1,\dots, x_k, \dots, x_n)$,   ${\bm t}=(t_1,\dots, t_k, \dots, t_n)$, ${\bm \theta}= (\theta_1, \dots, \theta_p)$ being 
$(x_k, t_k), 1 \leq k \leq n$  the pair $(x,t)$ where the $k-th$ data is taken, $\theta_i$, $\theta_j, 1 \leq i, j \leq p$ are the $i-th$ and $j-th$ parameters to be estimated  and
\begin{equation}
\label{Sensitivity}
S_{u}^{\theta_i}(x_k, t_k, \theta_1,...,\theta_N)=\dfrac{\partial u(x_k, t_k, \theta_1,...,\theta_N)}{\partial \theta_i},
\end{equation}
is the sensitivity of the temperature function  $u$, with respect to the parameter $\theta_i$.

In this work, we consider the ``D-optimal'' approach \cite{Banks10,Banks13,Banks15}. This is a simple and effective technique that consists of in minimizing the determinant of the inverse of the Fisher information matrix with respect to the data positions and instants, that is,  
\begin{equation}
\label{Inv_Fisher}
\minimo_{(\bm x, \bm t) \in A}  \left[\det\left(F^{-1}(\bm x, \bm t,\bm \theta)\right)\right],
\end{equation}
where $A$ is an admissible set which indicates where and when it is possible to take the data. 

By the definition given in \eqref{Fisher_n_parameters}, the minimization problem \eqref{Inv_Fisher} for $p=1$ reduces to find  $(x_k, t_k)$ where the sensitivity function takes the maximum absolute value.   Mathematically, 
\begin{equation}
\label{Max_sensi}
\maximo_{(x,t) \in [0,L] \times [0,\infty)} \left|S_{u}^{\alpha^2}(x,t,\alpha^2)\right|.
\end{equation}
%
%Hence, if the estimate is obtained from $n$ data, we look for $n$ pairs %$(x_i, t_j), \, j=1,...,n$ from \eqref{Max_sensi}. 

\section{Estimation of thermal diffusivity using D-Optimal Design}

Consider now the case of a bar of copper in Example \ref{Ex1} where the absolute sensitivity is calculated using the thermal diffusivity value for copper given in Table \ref{thermal_prop}. 
From the numerical simulations, we have observed that the maximum value is achieved at $(0.99 \, m, 21.0942 \, h) $ for which, the  absolute sensitivity is $\left| S_{u}^{\alpha^2}\right|=5.5586 \times 10^{10}.$

\begin{figure}[H]
\begin{center}
\includegraphics[width=0.42\textwidth]{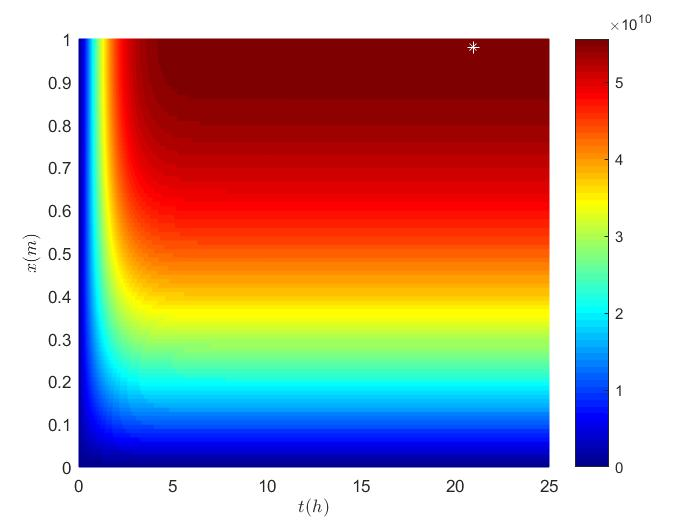}
\vspace{-0.5cm} 
\caption{$\left|S_{u}^{\alpha^2}(x,t,\alpha_{\rm{Cu}}^2)\right|$: Absolute  sensitivity function  for a copper bar.}
\vspace{-0.8cm} 
\label{Sensifunction}
\end{center}
\end{figure}
\begin{figure}[H]
\begin{center}
\includegraphics[width=0.42\textwidth]{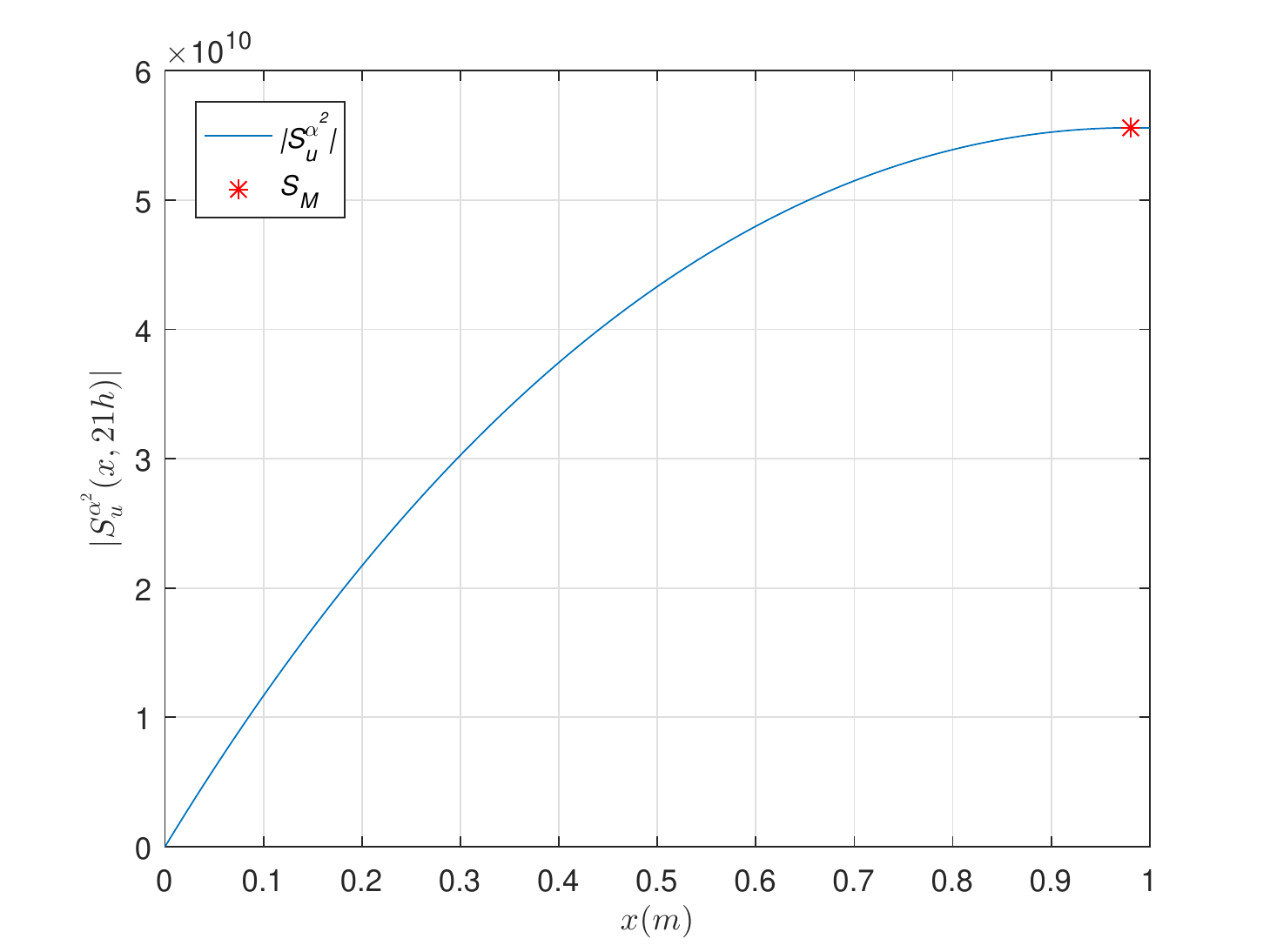}
\vspace{-0.5cm} 
\caption{$\left|S_{u}^{\alpha^2}(x,21h,\alpha_{\rm{Cu}}^2)\right|$: Absolute sensitivity function  at $t=21 \, h$  for a copper bar.}
%\vspace{-0.8cm} 
\label{Sensi_max}
\end{center}
\end{figure}
The absolute sensitivity function of the temperature with respect to the thermal diffusivity as a function of the space and time, is  shown in Figure \ref{Sensifunction}. The star symbol ``*'' indicates the position and instant at which the maximum value is achieved (white in Figure \ref{Sensifunction}, red in Figure \ref{Sensi_max}). 

Now, different positions on the bar are considered, and for each of them, we look for the instant that maximizes the absolute sensitivity. It is found that for all cases the absolute  sensitivity is  maximizes at about $t=21 \, h $ as shown in  Table \ref{table_max_sens}.

\begin{table}[h!]
\begin{center}
{\begin{tabular}{lccccc} \toprule
$$ x\,[m]$$ \quad \quad  & $$ \,\, t\,[s]$$ \quad \quad & $|S_{u}^{\alpha^2}| \,(\times 10^{10})$   \\ \midrule
\,\,\,0       &          0                           &                  0              \\        
      \,    $L/9$        &         21,0953           &                  1,2745         \\ 
            $2L/9$       &         21,0961           &                  2,3561         \\ 
            $3L/9$       &         21,0953           &                  3,2561         \\ 
	          $4L/9$       &         21,0930           &                  3,9939         \\ 					
				   $5L/9$       &         21,0978           &                  4,5792         \\ 
						$6L/9$       &         21,0898           &                  5,0215         \\ 
						$7L/9$       &         21,0925           &                  5,3286         \\
						$8L/9$       &         21,0950           &                  5,5063         \\
       \,    $L$         &         21,0939           &                  5,5582         \\ \bottomrule                            
\end{tabular}}
\end{center}
\vspace{-0.3cm} 
\caption{Optimal times for  measurements at different positions on a copper bar.}
\label{table_max_sens}
\end{table}

A set of $n= 5$ noisy simulated data at  equally spaced positions and $t=21 \, h$, is used to determine the thermal diffusivity coefficient of copper, where the noise level in data is  $\epsilon=1 \,{}^\circ C$. 
The estimated parameter value  results $\widehat{\alpha^2}=1.1252 \times 10^{-4} \, m^{2}/s $ so that, the relative error is $ Err_{Cu}=5.8391 \times 10^{- 5}$, or, $0.0058 \% $. Comparing this result with the ones obtained in Example \ref{Ex1}  it can be noticed that a greater accuracy is obtained when using optimal design for $n=5$ equally spaced data are considered at $t=12\, h$ and $t=24 \,h$ (see  Tables \ref{table3}-\ref{table4}). Moreover, the result is better than for all the cases analyzed in Example \ref{Ex1} (see in Tables \ref{table1}-\ref{table4}).

\section{Conclusion}
\label{Conclusion} \vspace{-4pt}

This work focuses on  the estimation of the thermal diffusivity coefficient from noisy temperature data numerically simulated. The process is modeled by a parabolic equation with  Dirichlet and Robin-type boundary conditions.
Good estimations result from simulated data at a fixed instant of time ( $t=12\, h$, $t=24 \, h$) in different positions of the bar. An analysis on the number of observations data required for the estimation is conducted concluding that five observations could be enough in order to obtain a good estimation.
Finally, an optimal design technique is applied to find the best positions and instants for observations to obtain the minimum estimation error that are used for the parameter identification.  Numerical results show that D-optimal design technique improves the accuracy in the estimation of the parameter. 

\vspace{10pt} \noindent
{\bf Acknowledgements:}  This work was partially supported by SOARD/AFOSR through grant FA9550-18-1-0523.

\end{document}